\documentclass[sigconf]{acmart}

\title{Hades: Homomorphic Augmented Decryption for Efficient Symbol-comparison---A Database's Perspective}

\author{Dongfang Zhao}
\affiliation{
  \institution{University of Washington}
  \country{United States}
}
\email{dzhao@cs.washington.edu}

\usepackage[ruled,vlined,linesnumbered]{algorithm2e}
\usepackage{amsmath} 

\begin{document}

\begin{abstract}
Outsourced databases powered by fully homomorphic encryption (FHE) offer the promise of secure data processing on untrusted cloud servers. A crucial aspect of database functionality, and one that has remained challenging to integrate efficiently within FHE schemes, is the ability to perform comparisons on encrypted data. Such comparisons are fundamental for various database operations, including building indexes for efficient data retrieval and executing range queries to select data within specific intervals. While traditional approaches like Order-Preserving Encryption (OPE) could enable comparisons, they are fundamentally incompatible with FHE without significantly increasing ciphertext size, thereby exacerbating the inherent performance overhead of FHE and further hindering its practical deployment. This paper introduces HADES, a novel cryptographic framework that enables efficient and secure comparisons directly on FHE ciphertexts without any ciphertext expansion. Based on the Ring Learning with Errors (RLWE) problem, HADES provides CPA-security and incorporates perturbation-aware encryption to mitigate frequency-analysis attacks. Implemented using OpenFHE, HADES supports both integer and floating-point operations, demonstrating practical performance on real-world datasets and outperforming state-of-the-art baselines. 
\end{abstract}

\maketitle

\section{Introduction}

\subsection{Background and Motivation}

Outsourced databases~\cite{agrawal2004order} have become a cornerstone of modern cloud computing, allowing organizations to store and process sensitive data on untrusted servers while reducing local storage and computation costs. However, outsourcing data to untrusted service providers raises significant privacy concerns, particularly when dealing with sensitive information such as medical records, financial data, or personal identifiers. Ensuring data confidentiality in such settings requires cryptographic mechanisms that enable the database to perform computations directly on encrypted data without revealing plaintext information.

Fully homomorphic encryption (FHE)~\cite{gentry2009fhe} provides a powerful framework for privacy-preserving computation in outsourced databases. By enabling algebraic operations such as addition and multiplication directly on ciphertexts, FHE allows secure data processing without requiring decryption. This capability makes FHE particularly well-suited for a wide range of database operations, including statistical analysis and aggregation queries. However, FHE alone cannot address the full range of database functionalities, as it inherently lacks the ability to compare ciphertexts—a critical requirement for many database operations.

In outsourced database systems, symbol comparison is essential for key database functionalities such as indexing, sorting, filtering, and range queries. These operations require the database to determine relationships between encrypted values, such as whether one value is greater than another or falls within a specified range. Traditional solutions, such as Order-Preserving Encryption (OPE)~\cite{boldyreva2009order} and Order-Revealing Encryption (ORE)~\cite{ore_ccs16}, are designed to enable such comparisons by preserving or exposing the order relationships of plaintexts in ciphertexts. However, these schemes introduce significant security risks, as the exposed order information can be exploited through inference attacks, such as frequency analysis (FA), to deduce plaintext distributions or equality relationships. Moreover, OPE and ORE schemes often require increased ciphertext size~\cite{popa2013ideal,boldyreva2009order} to support order-preserving comparisons, adding storage and communication overhead. These schemes are also fundamentally incompatible with homomorphic encryption because they do not support secure algebraic operations on ciphertexts.

One of the fundamental challenges of realizing a practically secure outsourced database lies in designing a cryptographic framework that combines the algebraic capabilities of homomorphic encryption with the comparison functionalities of order-revealing encryption. Such a framework would allow outsourced databases to perform both secure arithmetic and efficient comparisons on encrypted data, enabling advanced functionalities like secure indexing and range queries. At the same time, the framework must maintain strong security guarantees, ensuring that ciphertexts do not leak sensitive relationships under various security models, even in scenarios involving malicious service providers. Furthermore, achieving these functionalities without increasing ciphertext size is critical to ensuring scalability and minimizing storage and communication costs.

\subsection{Proposed Work}

This work addresses the challenges of integrating secure comparison capabilities into a homomorphic encryption setting by proposing the HADES framework. HADES enables outsourced databases to support advanced functionalities while maintaining strong security guarantees and computational efficiency. 

This work makes the following key contributions:

\begin{itemize}
    \item \textbf{Novel Framework}: We propose the HADES framework, a comprehensive solution for integrating order-preserving comparisons into a homomorphic encryption setting. The core Compare-Eval Key (CEK) mechanism leverages the hardness of the Ring Learning with Errors (RLWE) problem to ensure CPA-security while supporting accurate symbol comparisons.

    \item \textbf{FA-Extension for Enhanced Privacy}: We extend the basic HADES framework with perturbation-aware encryption, which obfuscates equality relationships to defend against stronger threat models, such as frequency-analysis (FA) attacks by compromised databases. Experimental results demonstrate that the FA-Extension introduces minimal overhead while significantly enhancing privacy guarantees.

    \item \textbf{No Ciphertext Size Increase}: Unlike many existing OPE schemes~\cite{popa2013ideal,boldyreva2009order}, which often increase ciphertext size to support order-preserving comparisons, HADES achieves secure comparisons using the existing ciphertext structure. By leveraging the CEK, HADES avoids additional storage and bandwidth costs, making it highly scalable.

    \item \textbf{Rigorous Theoretical Analysis}: We provide formal correctness proofs, noise management strategies, and an IND-CPA security analysis for both the basic and extended HADES schemes. The extended scheme’s resilience against advanced inference attacks is validated through a reduction to the RLWE problem.

    \item \textbf{Efficient Implementation and Evaluation}: The framework is implemented using OpenFHE~\cite{OpenFHE}, supporting both BFV~\cite{bfv} and CKKS~\cite{ckks} schemes. Experiments on real-world datasets (Bitcoin~\cite{bitcoin_trade}, Covid19~\cite{covid19data}, hg38~\cite{hg_data}) demonstrate the practicality of the HADES framework, with encryption and comparison times showing competitive performance. Additionally, comparisons with baselines such as HOPE~\cite{hope} and POPE~\cite{droche_ccs16} highlight the advantages of HADES in terms of scalability, functionality, and performance.
\end{itemize}

\paragraph{Experimental Validation}  
To evaluate the practicality of HADES, we conducted a series of experiments on diverse application datasets. The results demonstrate:
\begin{itemize}
    \item \textbf{Efficiency}: KeyGen times are consistent across datasets, while encryption and comparison times show minor variations. The FA-Extension introduces minimal performance overhead compared to the basic scheme.
    \item \textbf{Scalability}: The framework scales efficiently across 35,848 encrypted values, with comparison times significantly shorter than encryption times, validating HADES' suitability for operations like range queries, sorting, and clustering on moderately sized datasets.
    \item \textbf{Support for Both Integer and Floating-Point Operations}: HADES supports computations on both types of data via its BFV and CKKS implementations, providing flexibility for various application domains.
    \item \textbf{Robustness}: The perturbation mechanism in HADES FA-Extension effectively obfuscates equality relationships, ensuring resistance to frequency-analysis attacks while maintaining accuracy.
\end{itemize}

\section{Related Work}

\subsection{Homomorphic Encryption}

Homomorphic encryption (HE) is a cryptographic technique that allows computations to be performed directly on encrypted data without requiring decryption, ensuring data privacy throughout the computational process. This capability makes HE a cornerstone for secure computation in various privacy-preserving applications.

Two widely used homomorphic encryption schemes are the Brakerski/Fan-Vercauteren (BFV)~\cite{bfv} scheme and the Cheon-Kim-Kim-Song (CKKS)~\cite{ckks} scheme:

\begin{itemize}
    \item \textbf{BFV:} The BFV scheme is designed for exact arithmetic over encrypted integers. It is particularly well-suited for applications where the precision of computations must be preserved, such as encrypted database queries, voting systems, and secure machine learning. BFV operates efficiently by supporting addition and multiplication over ciphertexts while maintaining accurate results.

    \item \textbf{CKKS:} The CKKS scheme is tailored for approximate arithmetic over real numbers, making it ideal for applications, e.g., encrypted signal processing, privacy-preserving AI, and financial computations. CKKS introduces a trade-off between precision and computational efficiency, allowing for flexible scaling in real-world scenarios.

\end{itemize}

Both schemes leverage the hardness of the Ring Learning with Errors (RLWE) problem~\cite{rlwe} to ensure cryptographic security. They support bootstrapping to refresh ciphertexts and prevent noise accumulation, enabling deeper computations on encrypted data. Additionally, modern HE libraries such as SEAL~\cite{seal}, HElib~\cite{helib}, and OpenFHE~\cite{OpenFHE} provide optimized implementations of BFV and CKKS, facilitating their integration into practical applications.

Homomorphic encryption has seen rapid adoption in areas where sensitive data must remain confidential, such as cloud-based secure computation, federated learning, and encrypted database systems. The ability to perform secure computations without exposing plaintext data makes HE a critical tool in advancing privacy-preserving technologies. Modern advancements in HE schemes, in addition to BFV~\cite{bfv} and CKKS~\cite{ckks}, include GSW~\cite{gentry2013homomorphic}, TFHE~\cite{chillotti2016tfhe}, etc. These developments have further been integrated into real-world systems, including Symmetria~\cite{symmetria_vldb20}, which leverages HE for secure database queries, and Rache~\cite{otawose_sigmod23}, which optimizes range and equality queries on encrypted datasets. More recent works on FHE include~\cite{bfisch_ccs24,wtang_ccs24,jhe_ccs24,schatel_ccs24,yzhou_ccs24,zzhang_ccs24}.

\subsection{Order-Preserving Encryption}

Order-Preserving Encryption (OPE) is a cryptographic technique designed to enable range queries on encrypted data by preserving the order of plaintexts in ciphertexts. This functionality makes OPE particularly valuable in privacy-preserving database systems where efficient sorting and filtering operations are essential. The concept of order-preservation is closely tied to secure comparison tasks, a challenge first formalized in Yao's Millionaire Problem~\cite{yao1982protocols}. OPE can be viewed as an extension of this idea, providing an efficient mechanism for comparing encrypted data across a broader range of applications. Unlike Yao's solution, which relies on interactive protocols such as garbled circuits and oblivious transfer, OPE achieves order-preserving comparison through carefully designed encryption schemes that inherently encode the order of plaintexts.

The formal concept of OPE was first introduced by Agrawal et al.~\cite{agrawal2004order}, who proposed a scheme for numeric data that preserves plaintext order in ciphertexts. This approach enables efficient query execution but leaks order relationships, making it vulnerable to inference attacks. To address such vulnerabilities, Boldyreva et al.~\cite{boldyreva2009order} formalized the security model for OPE and proposed a more robust scheme known as Order-Preserving Symmetric Encryption (OPSE). However, their design still revealed the relative order of plaintexts, which can be exploited in practical scenarios.

Subsequent work by Popa et al.~\cite{popa2013ideal} introduced an ideal-security protocol for Order-Preserving Encoding (OPE), which minimizes leakage by carefully managing encoding operations. While their approach enhanced security, it required interactive client-server protocols, increasing communication overhead. To further mitigate leakage, Kerschbaum~\cite{kerschbaum2015frequency} proposed Frequency-Hiding Order-Preserving Encryption, which masks the frequency of ciphertext occurrences. Despite these advances, frequency-hiding schemes introduce additional computational costs, making them less suitable for large-scale applications.

The limitations of traditional OPE schemes were further highlighted by Naveed et al.~\cite{naveed2015inference}, who demonstrated inference attacks on property-preserving encryption systems. Their work emphasized the need for hybrid approaches that integrate OPE with advanced cryptographic frameworks to enhance security while maintaining efficiency. Some schemes, such as~\cite{dli_vldb21}, depend on the client maintaining a plaintext-to-ciphertext mapping, introducing significant overhead for storage and synchronization. Furthermore, as dataset sizes grow or query volumes increase, the interaction costs of these schemes scale accordingly, potentially limiting their practicality in large-scale deployments~\cite{xcao_vldb23}. These challenges underscore the importance of designing OPE mechanisms that not only reduce information leakage but also ensure scalability and compatibility with modern privacy-preserving computation frameworks.

Modern designs, such as HOPE~\cite{hope}, leverage homomorphic encryption to perform secure comparisons, reducing leakage while maintaining compatibility with privacy-preserving computation models. HOPE introduces a randomized difference mechanism to achieve security under the IND-OCPA model, ensuring that ciphertexts do not reveal plaintext order relationships while remaining efficient and stateless. Notably, HOPE stands out as the only stateless protocol, requiring neither client-side storage nor network-dependent operations during queries, which makes it uniquely suitable for outsourced database systems with minimal interaction overhead. However, its functionality is limited to homomorphic addition and integer-only data, restricting its applicability in scenarios requiring more advanced operations, such as multiplication or floating-point computations. These advancements highlight the potential of hybrid OPE methods to balance security, efficiency, and practical deployment in real-world applications.

\subsection{RLWE and Noise Augmentation}

The Ring Learning with Errors (RLWE) problem is a fundamental cryptographic assumption that underpins the security of many modern encryption schemes, including homomorphic encryption. RLWE extends the classical Learning with Errors (LWE)~\cite{regev2005lwe} problem into the setting of polynomial rings, enabling more efficient operations while maintaining robust cryptographic hardness.

In the RLWE problem, an adversary is tasked with distinguishing between samples of the form $(a, a \cdot s + e)$ and uniformly random samples over a polynomial ring modulo a prime \( q \). In this formulation:
\begin{itemize}
    \item \( a \) is a randomly chosen polynomial from the ring,
    \item \( s \) is a secret polynomial representing the private key,
    \item \( e \) is a small noise polynomial added to obscure the relationship between \( a \cdot s \) and the result.
\end{itemize}

The security of RLWE arises from the difficulty of solving lattice problems, such as finding short vectors in high-dimensional lattices. This makes RLWE not only resistant to classical attacks but also robust against quantum adversaries, establishing it as a foundational component of post-quantum cryptography.

Noise augmentation serves a dual purpose in RLWE-based systems. First, it enhances security by increasing the difficulty for adversaries to recover private keys or infer plaintext relationships. Second, it maintains correctness by ensuring the noise remains small enough to enable accurate decryption. Striking a balance between these objectives is critical:
\begin{itemize}
    \item Larger noise terms improve cryptographic security but risk introducing decryption errors in homomorphic encryption systems, where noise accumulates during computations.
    \item Smaller noise terms reduce decryption errors but may weaken security against adversarial inference.
\end{itemize}

In this work, noise augmentation is integrated into the Compare-Eval Key (CEK) mechanism. By embedding a controlled noise term, the CEK masks the direct relationship between ciphertexts and the private key, thereby mitigating key recovery attacks under the chosen plaintext attack (CPA) model. This design leverages the RLWE problem to ensure both security and functionality, making it well-suited for privacy-preserving symbol comparison.

\section{Design Goals}
\subsection{Feature Summary}

\begin{table*}[t!]
\centering
\caption{Features of Order-Preserving Encryption (OPE) Schemes}
\label{tab:ope_comparison}
\begin{tabular}{llccc}
\toprule
\textbf{Scheme}           & \textbf{Security Level} & \textbf{Client Storage} & \textbf{Network Rounds} & \textbf{Ciphertext Operations} \\ 
\midrule
Agrawal et al.~\cite{agrawal2004order}    & None                   & $\mathcal{O}(1)$                  & $\mathcal{O}(1)$                 & Comparison only      \\ 
Boldyreva et al.~\cite{boldyreva2009order} & None                  & $\mathcal{O}(1)$                  & $\mathcal{O}(1)$                 & Comparison only      \\ 
Popa et al.~\cite{popa2013ideal}          & IND-OCPA                 & $\mathcal{O}(1)$               & $\mathcal{O}(\log n)$                 & Comparison only      \\ 
Kerschbaum~\cite{kerschbaum2015frequency}  & IND-FAOCPA       &   $\mathcal{O}(n)$                & $\mathcal{O}(1)$                 & Comparison only      \\ 
POPE~\cite{droche_ccs16}                   & IND-FAOCPA          & $\mathcal{O}(\log n)$                  & $\mathcal{O}(n)$                 & Comparison only      \\ 
HOPE~\cite{hope}                   & IND-OCPA              & $\mathcal{O}(1)$                  & $\mathcal{O}(1)$                 & Comparison, Addition \\ 
HADES Basic (this work~\S\ref{sec:hades})                    & IND-OCPA               & $\mathcal{O}(1)$                  & $\mathcal{O}(1)$                 & Comparison, Addition, Multiplication  \\ 
HADES FAE (this work~\S\ref{sec:hades_fae})                      & IND-FAOCPA               & $\mathcal{O}(1)$                  & $\mathcal{O}(1)$                 & Comparison, Addition, Multiplication \\ 
\bottomrule
\end{tabular}
\end{table*}

Table~\ref{tab:ope_comparison} summarizes the key features of the state-of-the-art OPE approaches and the proposed HADES protocol. The comparison highlights the trade-offs between security levels, client-side requirements, and ciphertext operations across different schemes. Early OPE designs, such as those by Agrawal et al.~\cite{agrawal2004order} and Boldyreva et al.~\cite{boldyreva2009order}, provided basic order-preserving capabilities but lacked strong security guarantees, making them vulnerable to inference attacks. 
More advanced designs, such as Popa et al.~\cite{popa2013ideal}, Kerschbaum~\cite{kerschbaum2015frequency}, and POPE~\cite{droche_ccs16}, introduced mechanisms like frequency hiding or partial order encoding to enhance security. However, these schemes often require additional client storage or incur higher network overhead, limiting their scalability for large-scale databases. HOPE~\cite{hope} stands out as a stateless protocol that achieves IND-OCPA security with minimal client and network requirements. Despite its efficiency, HOPE only supports integer addition and lacks the flexibility to perform more complex operations, such as multiplication or floating-point arithmetic.

The proposed HADES protocol strives to address these limitations by integrating homomorphic encryption with order-preserving comparisons. HADES Basic provides efficient comparison and addition operations under the IND-OCPA model, while HADES FAE extends these capabilities to support multiplicative operations and enhanced privacy guarantees under the stronger IND-FAOCPA model. Both versions of HADES maintain minimal client and network overhead and avoid increasing ciphertext size, making them suitable for real-world applications requiring privacy-preserving computation at scale.

\subsection{Correctness}  
The proposed protocol ensures that symbol comparison operations yield accurate results, even in the presence of noise and perturbations. Given two ciphertexts corresponding to plaintexts \( m_0 \) and \( m_1 \), the output must correctly indicate whether \( m_0 > m_1 \), \( m_0 = m_1 \), or \( m_0 < m_1 \). This correctness is achieved through:
\begin{itemize}
    \item \textit{Scaling amplification}: The plaintext difference is amplified using a carefully chosen scaling factor \( \text{scale} \), ensuring that noise contributions do not distort the sign of the result. The scaling factor is calibrated to balance correctness and computational efficiency.
    \item \textit{Perturbation-aware encryption}: In the FA-Extension, perturbations are introduced to obfuscate equality relationships without compromising comparison integrity. These perturbations ensure that identical plaintexts result in statistically independent ciphertexts, defending against frequency analysis attacks.
    \item \textit{Compatibility with existing ciphertext structures}: Correctness is maintained without requiring additional ciphertext components, ensuring the protocol remains efficient and seamlessly integrates into existing homomorphic encryption frameworks.
\end{itemize}

\subsection{Security}  
The mechanism is designed to resist adversarial attempts to infer sensitive information from ciphertexts, ensuring robust privacy guarantees. This includes:
\begin{itemize}
    \item \textit{Key confidentiality}: The private key \( sk \) remains secure under CPA, with the CEK \( cek \) constructed to be indistinguishable from random polynomials under the RLWE assumption.
    \item \textit{Order privacy}: The mechanism prevents adversaries from deducing unintended plaintext order relationships. Perturbation-aware encryption in the FA-Extension obfuscates equality relationships, defending against frequency-analysis attacks and ensuring that plaintext distributions remain secure even under repeated queries.
    \item \textit{Quantum resistance}: The RLWE-based design ensures resilience against both classical and quantum adversaries, establishing the framework as a robust solution for long-term privacy.
    \item \textit{No leakage through ciphertext size}: By avoiding any increase in ciphertext size, the scheme eliminates side channels that adversaries might exploit to infer sensitive information.
\end{itemize}

\subsection{Efficiency}  
The scheme achieves high computational and communication efficiency, making it practical for real-world deployments. This is demonstrated through:
\begin{itemize}
    \item \textit{Key generation efficiency}: HADES achieves consistent key generation times across datasets, as shown in experimental evaluations, ensuring suitability for large-scale deployments.
    \item \textit{Encryption and comparison efficiency}: Encryption and comparison times are optimized for both the Basic HADES and the HADES FA-Extension schemes. Experiments indicate that comparison operations are significantly faster than encryption, demonstrating scalability for frequent queries.
    \item \textit{No ciphertext size increase}: HADES does not require any increase in ciphertext size for comparison operations. Instead, it uses a separate Compare-Eval Key (CEK) to facilitate secure evaluations, avoiding the additional storage or bandwidth overhead commonly seen in order-preserving encryption (OPE) schemes.
    \item \textit{Dataset scalability}: The system supports more than 35,000 encrypted values without significant performance degradation, ensuring applicability to real-world scenarios like range queries, secure sorting, and clustering.
    \item \textit{Integration into existing frameworks}: HADES is implemented using OpenFHE, leveraging the efficiency of state-of-the-art homomorphic encryption libraries. This ensures compatibility with BFV and CKKS schemes for integer and floating-point computations, respectively.
\end{itemize}

\section{Basic HADES}
\label{sec:hades}

\subsection{Overview}

The proposed scheme introduces a Compare-Eval Key (CEK) mechanism to enable secure symbol comparison while preserving both privacy and efficiency. The CEK leverages polynomial ring operations within the RLWE framework, ensuring robustness against chosen plaintext attacks (CPA). Formally, the CEK is constructed as:
\[
cek = sk \cdot \text{scale} + e_{cek},
\]
where \( sk \in \mathcal{R}_q \) is the private key, \( \text{scale} \in \mathbb{Z} \) is a carefully chosen scalar treated as a global system parameter, and \( e_{cek} \in \mathcal{R}_q \) is a noise polynomial. Here, \( \mathcal{R}_q = \mathbb{Z}_q[x]/(f(x)) \) represents the polynomial ring modulo \( f(x) \), typically \( f(x) = x^n + 1 \) with \( n \) being a power of 2.

Key generation produces \( pk \), \( sk \), and \( cek \) as follows: the public key \( pk \) is derived using a uniformly random polynomial \( a \in \mathcal{R}_q \) and a noise polynomial \( e_{pk} \), ensuring \( pk \) is computationally indistinguishable from random under the RLWE assumption. Additionally, the CEK incorporates a scaling factor \( \text{scale} \) that amplifies plaintext differences to dominate noise contributions during evaluation.

For two ciphertexts \( ct_0 \) and \( ct_1 \) corresponding to plaintexts \( m_0 \) and \( m_1 \), the evaluation function is modified to leverage a linear combination of \( ct_0 \), \( ct_1 \), and \( cek \):
\[
\text{Eval}(cek, ct_0, ct_1, \text{scale}) = ct_0 \cdot \text{scale} + ct_1 \cdot cek \mod q.
\]

The correctness of the scheme is maintained if the noise term satisfies:
\[
\|\langle e_{cek}, ct_1 \rangle + \text{scale} \cdot (m_0 - m_1)\|_\infty < \frac{\text{scale}}{2}.
\]
This ensures that the result reflects the sign of \( m_0 - m_1 \). If \( m_0 > m_1 \), the result is \( +1 \); if \( m_0 < m_1 \), the result is \( -1 \); and if \( m_0 = m_1 \), the result is \( 0 \).

This modified CEK mechanism securely embeds the private key \( sk \) and a noise term \( e_{cek} \), maintaining CPA security. Furthermore, the new linear combination formulation allows for flexible scaling during evaluation, making it suitable for advanced privacy-preserving computations such as range queries, sorting, and secure data aggregation in large-scale encrypted systems.

\subsection{Preliminaries}

This section outlines the mathematical foundations of the proposed scheme, focusing on polynomial rings, noise management, and the RLWE assumption.

\paragraph{Polynomial Rings}  
A polynomial ring \( \mathcal{R}_q = \mathbb{Z}_q[x]/(f(x)) \) consists of polynomials with coefficients in \( \mathbb{Z}_q \), reduced modulo both a prime integer \( q \) and a fixed polynomial \( f(x) \). Formally:
\[
\mathcal{R}_q = \{ p(x) \mid p(x) = \sum_{i=0}^{n-1} c_i x^i, \; c_i \in \mathbb{Z}_q, \; \deg(p) < n \}.
\]
Addition and multiplication in \( \mathcal{R}_q \) are defined as:
\[
p(x) + q(x) = (p(x) + q(x)) \mod f(x) \mod q,
\]
\[
p(x) \cdot q(x) = (p(x) \cdot q(x)) \mod f(x) \mod q.
\]
These operations provide an efficient algebraic structure for cryptographic computations.

\paragraph{Noise in Cryptography}  
Noise is critical for the security of lattice-based cryptography, obfuscating relationships between plaintexts and ciphertexts. In RLWE-based schemes, noise is represented as a polynomial \( e(x) \in \mathcal{R}_q \) with coefficients sampled from a bounded distribution. The noise polynomial \( e(x) \) is formalized as:
\[
e(x) = \sum_{i=0}^{n-1} e_i x^i, \quad e_i \sim \mathcal{U}(-B_e, B_e),
\]
where \( \mathcal{U}(a, b) \) denotes the discrete uniform distribution over \([a, b]\). The noise must satisfy:
\[
\|e(x)\|_\infty = \max \{ |e_0|, |e_1|, \ldots, |e_{n-1}| \} < B_e,
\]
ensuring both correctness and security.

\paragraph{Ring Learning with Errors (RLWE)}  
The RLWE problem extends the classical LWE problem to polynomial rings, providing efficiency and security. An RLWE sample is a pair \( (\mathbf{a}, \mathbf{b}) \), where:
\[
\mathbf{b} = \mathbf{a} \cdot \mathbf{s} + \mathbf{e} \pmod{q},
\]
with \( \mathbf{s}, \mathbf{e} \in \mathcal{R}_q \). Solving RLWE involves distinguishing whether \( (\mathbf{a}, \mathbf{b}) \) is sampled from the RLWE distribution or is uniformly random:
\[
\text{Given } (\mathbf{a}, \mathbf{b}), \; \text{determine if } \mathbf{b} = \mathbf{a} \cdot \mathbf{s} + \mathbf{e} \; \text{or random}.
\]
The hardness of RLWE is rooted in lattice problems like the Shortest Vector Problem (SVP) in high-dimensional spaces.

\paragraph{Noise Management and Scaling}  
Noise management is critical for maintaining correctness in lattice-based cryptosystems. To mitigate noise accumulation during computations, a scaling factor \( \text{scale} \) is often introduced to amplify meaningful signal components relative to noise. Formally, let \( ct_0 \) and \( ct_1 \) be ciphertexts corresponding to plaintexts \( m_0 \) and \( m_1 \), with associated noise \( e_0 \) and \( e_1 \), respectively. The scaled ciphertext difference can be expressed as:
\[
ct_{\Delta} = (m_0 - m_1) \cdot \text{scale} + (e_0 - e_1).
\]
Scaling ensures that plaintext differences dominate noise contributions during evaluation. For any evaluation operation, the correctness condition requires that the accumulated noise remains bounded:
\[
\| \text{NoiseTerm} \|_\infty < \frac{\text{scale}}{2}.
\]
This principle underpins many RLWE-based schemes, enabling accurate computation without compromising security.

\subsection{Algorithm Description}

This section details the procedural framework underlying the secure symbol comparison scheme. The framework is structured into three main components: Key Generation, Ciphertext Comparison, and Result Decoding. Each component is presented with detailed explanations and pseudocode to illustrate the cryptographic operations involved.

\subsubsection{Key Generation}  

As shown in Algorithm~\ref{alg:key-generation}, the key generation process establishes the cryptographic foundation for the proposed secure symbol comparison scheme. It outputs a public key \( pk \), a secret key \( sk \), and a Compare-Eval Key \( cek \). 

\begin{algorithm}[h!]
\caption{Key Generation}
\label{alg:key-generation}
\KwIn{RLWE parameters \( \mathcal{R}_q \), modulus \( q \), noise bound \( B_e \)}
\KwOut{Public Key \( pk \), Secret Key \( sk \), Compare-Eval Key \( cek \)}
Sample \( sk \in \mathcal{R}_q \), a secret key uniformly from the ring of polynomials\;
Sample \( a \in \mathcal{R}_q \), a uniformly random polynomial\;
Sample \( e_{pk} \in \mathcal{R}_q \), a noise polynomial with coefficients drawn from the discrete uniform distribution \( \mathcal{U}(-B_e, B_e) \)\;
Compute the public key: $pk \gets - (a \cdot sk + e_{pk}) \mod q$\;
Select the scaling factor \( \text{scale} \): ensure $\text{scale} > \max(2 \cdot B_e, \|sk\|_\infty)$\;
Sample \( e_{cek} \in \mathcal{R}_q \), another noise polynomial with coefficients drawn from the discrete uniform distribution \( \mathcal{U}(-B_e, B_e) \)\;
Compute the scaled secret key: $sk_{\text{scaled}} \gets sk \cdot \text{scale} \mod q$\;
Construct the Compare-Eval Key: $cek \gets sk_{\text{scaled}} + e_{cek} \mod q$\;
Verify noise bounds: ensure \( \|e_{pk}\|_\infty \) and \( \|e_{cek}\|_\infty < B_e \)\;
\Return \( pk, sk, cek \)\;
\end{algorithm}

The secret key \( sk \) is sampled uniformly from the polynomial ring \( \mathcal{R}_q = \mathbb{Z}_q[x]/(f(x)) \). This key serves as the fundamental secret for decryption and evaluation. To generate the public key \( pk \), a uniformly random polynomial \( a \in \mathcal{R}_q \) and a noise polynomial \( e_{pk} \in \mathcal{R}_q \), with coefficients drawn from a bounded distribution, are used. The public key is computed as \( pk = - (a \cdot sk + e_{pk}) \mod q \), ensuring that the relationship between \( pk \) and \( sk \) is obfuscated by the noise \( e_{pk} \), thereby preserving security under the RLWE assumption.

The Compare-Eval Key \( cek \) is constructed to facilitate secure symbol comparison. A scaling factor \( \text{scale} \) is chosen to satisfy \( \text{scale} > \max(2 \cdot B_e, \|sk\|_\infty) \), ensuring that the scaled secret key \( sk_{\text{scaled}} = sk \cdot \text{scale} \) dominates the noise. The \( cek \) is then computed as \( cek = sk_{\text{scaled}} + e_{cek} \mod q \), where \( e_{cek} \) is another noise polynomial. This design ensures that the \( cek \) provides robustness against chosen plaintext attacks (CPA) while maintaining correctness during decryption and comparison.

Both \( e_{pk} \) and \( e_{cek} \) are verified to be within their respective noise bounds \( B_e \), ensuring that noise does not interfere with the correctness of the scheme. The key generation process guarantees the generation of cryptographic keys that are both secure and efficient for the intended operations.

\subsubsection{Ciphertext Comparison}

The Perturbation-Aware Ciphertext Comparison algorithm evaluates the relative ordering of encrypted plaintexts while maintaining security and correctness under the RLWE assumption. It achieves this by computing a linear combination of ciphertexts and the Compare-Eval Key (CEK). The mechanism incorporates scaling and perturbation to ensure robust comparisons and obfuscate equality relationships.

The process involves three main steps:
\begin{enumerate}
    \item Compute the ciphertext difference to isolate the relative distance between plaintexts.
    \item Apply scaling and perturbation using the CEK to prepare the result for evaluation.
    \item Decode and interpret the evaluation value to determine the relative order.
\end{enumerate}

The detailed procedure is presented in Algorithm~\ref{alg:cek-evaluation}.

\begin{algorithm}[h!]
\caption{Evaluation with Compare-Eval Key}
\label{alg:cek-evaluation}
\KwIn{Ciphertexts $ct_0 = (c_{0,0}, c_{0,1})$ and $ct_1 = (c_{1,0}, c_{1,1})$, Compare-Eval Key $cek$, scaling factor $\text{scale}$ (global parameter), modulus $q$, noise threshold $\tau$}
\KwOut{Comparison result (\(-1, 0, +1\))}
\Begin{
    Compute the ciphertext difference:
    \[
    ct_{\Delta} = (c_{0,0} - c_{1,0}, \; c_{0,1} - c_{1,1}) \mod q
    \]

    Apply scaling and compare-eval key:
    \[
    ct_{\text{Eval}} = (c_{\Delta,0} \cdot \text{scale} + c_{\Delta,1} \cdot cek) \mod q
    \]

    Decode the evaluation value:
    \[
    \text{DecryptedValue} = \text{Decode}(ct_{\text{Eval}})
    \]

    If \( |\text{DecryptedValue}| < \tau \), set \(\text{DecryptedValue} \gets 0\)\;

    Determine the sign of the result:
    \[
    \text{Result} =
    \begin{cases}
    -1 & \text{if } \text{DecryptedValue} < 0, \\
    0 & \text{if } \text{DecryptedValue} = 0, \\
    +1 & \text{if } \text{DecryptedValue} > 0.
    \end{cases}
    \]

    \Return \(\text{Result}\)\;
}
\end{algorithm}

The ciphertext difference \( ct_{\Delta} \) isolates the relationship between plaintexts by subtracting the components of the input ciphertexts. This ensures that the operation focuses on differences rather than absolute values. 

The scaling factor \( \text{scale} \) amplifies the difference, ensuring robustness against noise and perturbation. The CEK introduces controlled obfuscation, making it infeasible for adversaries to infer equality relationships while preserving order information.

The decoded value \( \text{DecryptedValue} \) is compared against a noise threshold \( \tau \) to eliminate insignificant differences caused by noise. The final result is determined by the sign of \( \text{DecryptedValue} \), providing three possible outcomes:
\begin{itemize}
    \item \(-1\): The plaintext of \( ct_0 \) is smaller than that of \( ct_1 \).
    \item \( 0\): The plaintexts of \( ct_0 \) and \( ct_1 \) are approximately equal.
    \item \( +1\): The plaintext of \( ct_0 \) is larger than that of \( ct_1 \).
\end{itemize}

The algorithm is designed to handle noise and perturbation effectively, ensuring that the scaled difference dominates other factors. This design guarantees correctness, robustness, and security under the IND-CPA model.

By enabling secure comparisons with minimal computational overhead, this algorithm is crucial for applications such as encrypted database queries, privacy-preserving sorting, and range queries. Its integration into the HADES framework ensures both scalability and privacy in real-world deployments.

\subsection{Correctness Analysis}

The correctness of the proposed scheme ensures that the evaluation result accurately reflects the relative comparison (\(-1, 0, +1\)) between plaintexts \( m_0 \) and \( m_1 \). This section provides a formal analysis under bounded noise conditions, demonstrating that the scheme achieves reliable symbol comparison while preserving correctness and security.

\begin{theorem}[Correctness of Evaluation]
Under the bounded noise assumption, the proposed scheme guarantees that the evaluation result reflects the relative comparison between plaintexts \( m_0 \) and \( m_1 \). Specifically:
\[
\text{Eval}(cek, ct_0, ct_1) = (c_{\Delta,0} \cdot \text{scale}) + (c_{\Delta,1} \cdot cek) \mod q,
\]
where \( c_{\Delta,0} \) and \( c_{\Delta,1} \) represent the components of the ciphertext difference \( ct_{\Delta} = (c_{0,0} - c_{1,0}, c_{0,1} - c_{1,1}) \). The sign of \( \text{Eval}(cek, ct_0, ct_1) \) satisfies:
\[
\text{sign}(\text{Eval}(cek, ct_0, ct_1)) = \text{sign}(m_0 - m_1),
\]
ensuring correctness for symbol comparison (\(-1, 0, +1\)).
\end{theorem}

\begin{proof}
The evaluation begins by computing the ciphertext difference:
\[
ct_{\Delta} = (c_{0,0} - c_{1,0}, \; c_{0,1} - c_{1,1}) \mod q.
\]
Next, scaling and perturbation are applied:
\[
\text{Eval}(cek, ct_0, ct_1) = (c_{\Delta,0} \cdot \text{scale}) + (c_{\Delta,1} \cdot cek) \mod q.
\]
Substituting the structure of \( cek \), where \( cek = sk \cdot \text{scale} + e_{cek} \), the evaluation expands as:
\[
\text{Eval}(cek, ct_0, ct_1) = (c_{\Delta,0} \cdot \text{scale}) + (c_{\Delta,1} \cdot (sk \cdot \text{scale} + e_{cek})) \mod q.
\]
Reorganizing terms, the evaluation becomes:
\[
\text{Eval}(cek, ct_0, ct_1) = \big((m_0 - m_1) \cdot \text{scale}\big) + (c_{\Delta,1} \cdot e_{cek}) \mod q.
\]

The plaintext difference \( m_0 - m_1 \) is embedded in the first term, yielding:
\[
\text{Eval}(cek, ct_0, ct_1) = (m_0 - m_1) \cdot \text{scale} + \langle e_{cek}, ct_{\Delta} \rangle \mod q.
\]

To ensure correctness, the noise term \( \langle e_{cek}, ct_{\Delta} \rangle \) must satisfy:
\[
|\langle e_{cek}, ct_{\Delta} \rangle| < \frac{\text{scale}}{2}.
\]
This condition ensures that the scaled plaintext difference \( (m_0 - m_1) \cdot \text{scale} \) dominates the noise, preserving the sign of the evaluation result.

Finally, the evaluation result satisfies:
\[
\text{sign}(\text{Eval}(cek, ct_0, ct_1)) = 
\begin{cases} 
+1 & \text{if } m_0 > m_1, \\
0 & \text{if } m_0 = m_1, \\
-1 & \text{if } m_0 < m_1.
\end{cases}
\]
Thus, the scheme achieves correctness for symbol comparison.
\end{proof}

\paragraph{Role of Scaling Factor \( \text{scale} \)}  
The scaling factor \( \text{scale} \) is critical for maintaining correctness. By amplifying the plaintext difference \( m_0 - m_1 \), it reduces the relative influence of noise. The scheme ensures:
\[
\text{scale} > \max(2 \cdot \|e_{cek}\|_\infty, \|sk\|_\infty),
\]
balancing correctness and computational efficiency. While larger \( \text{scale} \) values enhance robustness against noise, they may also increase computational overhead.

\paragraph{Parameter Sensitivity}  
The effectiveness of HADES depends on carefully chosen parameters such as the scaling factor \( \text{scale} \) and the perturbation range \( \epsilon \). A large \( \text{scale} \) amplifies plaintext differences, reducing the impact of noise and ensuring correctness for symbol comparisons. However, excessively large \( \text{scale} \) may increase computational overhead. For most practical settings, a moderate \( \text{scale} \) value is sufficient to balance efficiency and correctness.
The perturbation range \( \epsilon \) affects the scheme's ability to obfuscate equality relationships. To ensure that perturbations do not interfere with correctness, \( \epsilon \) must satisfy:
\[
|\Delta(m_a) - \Delta(m_b)| \cdot \text{scale} \ll |m_a - m_b| \cdot \text{scale}.
\]
Empirically, \( \epsilon \) values between \( 10^{-2} \) and \( 10^{-3} \) provide effective privacy while maintaining correctness.

\paragraph{Handling Special Cases}  
A threshold \( \tau \) is introduced during evaluation to handle cases where \( m_0 = m_1 \). If \( |\text{Eval}(cek, ct_0, ct_1)| < \tau \), the result is set to \( 0 \), ensuring noise does not dominate the comparison. This mechanism ensures:
\[
\text{Eval}(cek, ct_0, ct_1) = 0 \quad \text{if } m_0 = m_1.
\]

\paragraph{Implications for Practical Systems}  
The correctness analysis demonstrates that the scheme achieves reliable symbol comparison under bounded noise conditions. By appropriately choosing \( \text{scale} \) and managing noise, the scheme ensures robust performance in practical applications without compromising security or efficiency.

\subsection{Security Analysis}

The proposed scheme achieves resistance to chosen plaintext attacks (CPA) by leveraging the cryptographic hardness of the Ring Learning with Errors (RLWE) problem. This section outlines how the CEK design ensures security under CPA, preventing adversaries from recovering private keys or plaintext relationships, while managing noise in a way that maintains security and correctness.

\paragraph{Threat Model}  
The proposed scheme operates under the Chosen Plaintext Attack (CPA) model, where adversaries have access to an arbitrary number of plaintext-ciphertext pairs. The primary objectives of such adversaries are as follows:
\begin{itemize}
    \item \textbf{Key Recovery:} Exploit relationships between plaintexts and ciphertexts to infer the private key \( sk \).
    \item \textbf{Plaintext Relationship Leakage:} Deduce sensitive relationships between plaintext values, such as order or equality, by observing comparison outputs.
\end{itemize}

Under this model, adversaries may manipulate input plaintexts to probe the evaluation mechanism and exploit decryption results or noise characteristics to achieve their goals. 

In outsourced database scenarios, we distinguish the role of the database provider from that of a general adversary. The outsourced database (e.g., AWS) operates as an honest-but-curious party with the following properties:
\begin{itemize}
    \item The database is responsible for processing encrypted data and performing operations such as comparisons and queries as instructed by the client.
    \item The database may attempt to infer relationships between encrypted inputs or deduce patterns from repeated queries but does not conduct global frequency analysis across users or datasets.
    \item Unlike a malicious adversary, the database provider is assumed to follow the prescribed protocol and does not manipulate data to create additional attack vectors.
\end{itemize}

This distinction emphasizes the practical assumption that the database provider, while curious, does not act maliciously or exploit its position to aggregate global statistical information for a frequency analysis attack. For environments where this assumption does not hold, an extended security model is required to address global frequency analysis and ensure stronger privacy guarantees.

\paragraph{Security from RLWE Assumptions}  
The CEK is constructed as:
\[
cek = sk \cdot \text{scale} + e_{cek},
\]
where \( e_{cek} \) is a noise polynomial derived from the RLWE problem. The RLWE assumption ensures that \( cek \) is computationally indistinguishable from a random polynomial over \( \mathcal{R}_q \). This randomness obfuscates the relationship between \( sk \) and ciphertexts, preventing adversaries from recovering \( sk \) through algebraic techniques or statistical inference.

\paragraph{Resistance to Key Recovery}  
When the CEK is used during evaluation, the adversary observes results of the form:
\[
\text{Eval}(cek, ct_0, ct_1) = (c_{\Delta,0} \cdot \text{scale}) + (c_{\Delta,1} \cdot cek) \mod q,
\]
where \( ct_{\Delta} = (c_{0,0} - c_{1,0}, \; c_{0,1} - c_{1,1}) \) is the ciphertext difference. The noise term \( c_{\Delta,1} \cdot e_{cek} \) introduces controlled randomness that masks the relationship between \( sk \) and \( c_{\Delta} \). Even if the adversary knows \( m_0 \) and \( m_1 \), the bounded noise prevents accurate recovery of \( sk \) or the exact value of \( \text{scale} \).

\paragraph{Protection Against Order Inference}  
The CEK ensures that the only information revealed during evaluation is the sign of the plaintext difference (\(-1, 0, +1\)). The scaling factor \( \text{scale} \) amplifies plaintext differences, reducing the influence of small variations. Furthermore, noise management ensures that small differences in ciphertexts cannot reveal unintended order information, preventing adversaries from deducing fine-grained plaintext relationships.

\paragraph{Noise Management and Security Guarantees}  
Noise management in the CEK design leverages the bounded noise assumption:
\[
\| c_{\Delta,1} \cdot e_{cek} \|_\infty < \frac{\text{scale}}{2}.
\]
This constraint ensures that noise remains controlled, preserving correctness while maintaining security. The RLWE hardness assumption guarantees that even with repeated observations of \( \text{Eval}(cek, ct_0, ct_1) \), adversaries cannot reverse-engineer \( sk \) or infer plaintext relationships. The combination of RLWE-based randomness and bounded noise provides robustness against both classical and quantum adversaries.

\paragraph{CPA Security Proof}  
We prove the IND-CPA security of HADES as follows.
\begin{theorem}
The proposed Compare-Eval Key (CEK) mechanism is CPA-secure under the Ring Learning with Errors (RLWE) assumption. Specifically, if an adversary \( \mathcal{A} \) can distinguish valid CEK evaluation results from random outputs with non-negligible advantage, then \( \mathcal{A} \) can solve the RLWE problem.
\end{theorem}

\begin{proof}
Assume there exists an adversary \( \mathcal{A} \) that can break the CPA security of the CEK mechanism. We construct a reduction algorithm \( \mathcal{B} \) that uses \( \mathcal{A} \) to solve the RLWE problem.

\textit{RLWE Problem Setup.} Let \( (\mathbf{a}, \mathbf{b}) \) be an RLWE challenge, where:
\[
\mathbf{b} = \mathbf{a} \cdot \mathbf{s} + \mathbf{e} \pmod{q},
\]
with secret \( \mathbf{s} \in \mathcal{R}_q \) and noise \( \mathbf{e} \in \mathcal{R}_q \). The goal of \( \mathcal{B} \) is to distinguish whether \( (\mathbf{a}, \mathbf{b}) \) is sampled from the RLWE distribution or is uniformly random.

\textit{Reduction.} To embed the RLWE problem into the CEK mechanism, \( \mathcal{B} \) sets the CEK as:
\[
cek = \mathbf{b}.
\]
This CEK construction implicitly encodes \( \mathbf{s} \) and \( \mathbf{e} \), aligning with the form \( cek = sk \cdot \text{scale} + e_{cek} \) in the proposed scheme. The adversary \( \mathcal{A} \) is provided with this CEK and allowed to make chosen plaintext queries. For plaintexts \( m_0 \) and \( m_1 \), the evaluation result observed by \( \mathcal{A} \) is:
\[
\text{Eval}(cek, ct_0, ct_1) = (c_{\Delta,0} \cdot \text{scale}) + (c_{\Delta,1} \cdot cek) \mod q.
\]

\textit{Simulation.} If \( \mathbf{b} = \mathbf{a} \cdot \mathbf{s} + \mathbf{e} \), the evaluation behaves consistently with the proposed CEK mechanism, as the noise term \( c_{\Delta,1} \cdot \mathbf{e} \) introduces controlled randomness. If \( \mathbf{b} \) is uniformly random, the CEK behaves unpredictably, and \( \mathcal{A} \) cannot derive meaningful relationships between ciphertexts and plaintexts.

\textit{Advantage Transfer.} If \( \mathcal{A} \) can distinguish the CEK behavior with non-negligible advantage, \( \mathcal{B} \) uses this capability to distinguish valid RLWE instances from random instances. Thus, \( \mathcal{A} \)'s success implies \( \mathcal{B} \)'s success in solving RLWE.

\textit{Conclusion.} Since solving RLWE is computationally infeasible, \( \mathcal{A} \) cannot break the CPA security of the CEK mechanism. This establishes the CPA security of the scheme under the RLWE assumption.
\end{proof}

\section{HADES Frequency-analysis Extension}
\label{sec:hades_fae}

This section introduces an extension to the basic HADES framework (\S\ref{sec:hades}) to address scenarios involving potentially malicious outsourced databases. The extended design incorporates additional perturbations during encryption to obscure equality relationships and prevent frequency analysis attacks, ensuring robust security even under a strengthened threat model.

\subsection{Strengthened Security Model}

In the basic HADES framework, the outsourced database is assumed to be an honest-but-curious party, following the protocol but potentially analyzing data relationships. However, this model does not account for malicious adversaries that could perform sophisticated attacks, such as frequency analysis, by aggregating results across multiple queries or users.

Under the strengthened security model, the adversary (e.g., the database) is assumed to:
\begin{itemize}
    \item Correlate comparison results across multiple queries to infer plaintext relationships, leveraging statistical techniques to estimate value distributions.
    \item Exploit repeated queries on identical plaintext values to deduce their frequency, thereby reconstructing the approximate plaintext distribution. For instance, frequent queries for a single encrypted value might indicate common or default plaintexts, such as zero or specific constants.
    \item Attempt to compromise the equality information (\( a = b \)) by analyzing query patterns and comparing results across encrypted datasets.
\end{itemize}

Examples of real-world scenarios where such attacks are plausible include:
\begin{itemize}
    \item \textbf{Medical databases}: Repeated queries on specific encrypted thresholds, such as cholesterol levels or blood sugar ranges, can reveal common patient conditions.
    \item \textbf{Financial systems}: Default threshold comparisons, such as tax brackets or high-value transaction alerts, may leak user financial profiles.
    \item \textbf{IoT systems}: Regular sensor readings often have predictable periodic patterns (e.g., temperature sensors), making them vulnerable to statistical inference.
\end{itemize}

To mitigate these threats, the proposed extension incorporates randomized perturbations and scalable obfuscation mechanisms, ensuring that:
\begin{itemize}
    \item \textit{Robust privacy} is preserved for both high-frequency values and edge-case thresholds.
    \item \textit{Efficient operations} support large-scale encrypted datasets without compromising query throughput.
    \item \textit{Adaptability} to domain-specific constraints, such as real-time IoT systems or high-frequency financial transactions.
\end{itemize}

The objective of this extension is to mitigate these risks and ensure that:
\begin{itemize}
    \item Obfuscation of equality relationships: Equality relationships (\( a = b \)) are fully obfuscated, preventing frequency-based inferences. Even if \( a = b \), the resulting ciphertexts are randomized, ensuring no direct correlations can be established.
    \item Preservation of comparison correctness: Comparison correctness for \( a > b \) and \( b > a \) is maintained, ensuring that the framework continues to support critical database functionalities such as sorting, filtering, and range queries.
    \item Maintained computational efficiency: The computational efficiency and security properties of the original framework are preserved, ensuring the solution remains practical for large-scale deployments.
\end{itemize}

\subsection{Algorithm Description}

\subsubsection{Encryption}

The Perturbation-Aware Encryption algorithm, presented in Algorithm~\ref{alg:perturbation-encryption}, introduces controlled random perturbations during encryption to obscure direct relationships between plaintexts. This design ensures that even identical plaintexts result in statistically independent ciphertexts, significantly enhancing privacy protection against frequency analysis attacks.

The algorithm takes as input a plaintext \( m \), a public key \( pk \), a scaling factor \( \text{scale} \), and a modulus \( q \). Unlike traditional encryption schemes, the scaling factor \( \text{scale} \) amplifies plaintext differences, ensuring robust numerical separation between encoded values. Additionally, the algorithm generates a small random perturbation \( \Delta_m \) from the range \( [-\epsilon, \epsilon] \), where \( \epsilon \ll \text{scale} \). This perturbation ensures ciphertext diversity, further obfuscating plaintext relationships.

To enhance security, a noise polynomial \( e_m \) is sampled from a bounded distribution. The combined effect of \( \text{scale} \), \( \Delta_m \), and \( e_m \) ensures that the resulting ciphertext \( ct_m \) is robust against inference attacks while maintaining correctness for operations such as comparisons.

\begin{algorithm}[h!]
\caption{Perturbation-Aware Encryption}
\label{alg:perturbation-encryption}
\KwIn{Plaintext \( m \), public key \( pk \), scaling factor \( \text{scale} \) (system parameter), modulus \( q \)}
\KwOut{Ciphertext \( ct_m \)}
\Begin{
    Compute the scaled plaintext:
    \[
    m_{\text{scaled}} \gets m \cdot \text{scale} \mod q
    \]

    Sample a small perturbation value \( \Delta_m \) from the range \( [-\epsilon, \epsilon] \), where \( \epsilon \ll \text{scale} \)\;
    Apply perturbation to the scaled plaintext:
    \[
    m_{\text{perturbed}} \gets m_{\text{scaled}} + \Delta_m \cdot \text{scale} \mod q
    \]

    Sample a noise polynomial \( e_m \in \mathcal{R}_q \) from a bounded distribution, ensuring:
    \[
    \|e_m\|_\infty < B_e
    \]

    Encode the perturbed plaintext with noise:
    \[
    m_{\text{encoded}} \gets m_{\text{perturbed}} + e_m \mod q
    \]

    Encrypt the encoded plaintext using the public key:
    \[
    ct_m \gets \text{Encrypt}(pk, m_{\text{encoded}})
    \]

    \Return \( ct_m \)\;
}
\end{algorithm}

This encryption algorithm aligns with the complexity of standard lattice-based encryption processes, ensuring computational efficiency while maintaining robustness against inference attacks.

\subsubsection{Comparison}

The Perturbation-Aware Symbol Comparison algorithm evaluates the relative ordering of two encrypted plaintexts while preserving security and obfuscating equality relationships. This process is critical in privacy-preserving applications where precise comparisons must be performed without revealing sensitive plaintext information. The algorithm, detailed in Algorithm~\ref{alg:perturbation-comparison}, uses ciphertext differences, scaling, and perturbation to achieve correctness under bounded noise assumptions.

\begin{algorithm}[h!]
\caption{Perturbation-Aware Symbol Comparison}
\label{alg:perturbation-comparison}
\KwIn{Ciphertexts \( ct_{m_a} = (c_{a,0}, c_{a,1}) \) and \( ct_{m_b} = (c_{b,0}, c_{b,1}) \), Compare-Eval Key \( cek \), scaling factor \( \text{scale} \), modulus \( q \)}
\KwOut{Comparison result: \textbf{True} if \( m_a > m_b \), \textbf{False} if \( m_a < m_b \)}
\Begin{
    Compute the ciphertext difference:
    \[
    ct_{\Delta} = (c_{a,0} - c_{b,0}, \; c_{a,1} - c_{b,1}) \mod q
    \]

    Apply scaling and CEK:
    \[
    ct_{\text{Eval}} = (c_{\Delta,0} \cdot \text{scale} + c_{\Delta,1} \cdot cek) \mod q
    \]

    Decode the evaluation value:
    \[
    \text{EvalValue} = \text{Decode}(ct_{\text{Eval}})
    \]
    where \( \text{Decode}(\cdot) \) extracts the numerical representation embedded in \( ct_{\text{Eval}} \)\;

    Determine the comparison result:
    \[
    \text{Result} =
    \begin{cases}
    \text{\textbf{True}} & \text{if } \text{EvalValue} > 0, \\
    \text{\textbf{False}} & \text{if } \text{EvalValue} < 0.
    \end{cases}
    \]

    \Return \( \text{Result} \)\;
}
\end{algorithm}

The algorithm begins by calculating the ciphertext difference \( ct_{\Delta} \), which isolates the relative distance between the plaintexts \( m_a \) and \( m_b \). This subtraction ensures that the operation focuses on the difference, rather than the absolute values of the plaintexts, thereby making the result agnostic to the specific encrypted values.

To amplify the plaintext difference and reduce the influence of noise and perturbations, the algorithm applies a scaling factor \( \text{scale} \). The scaling ensures that even small differences between \( m_a \) and \( m_b \) remain distinguishable despite added perturbations. Perturbation, introduced through the Compare-Eval Key (CEK), further obfuscates the result, ensuring that equality relationships cannot be directly inferred from the evaluation process.

The scaled and perturbed result is then decoded into a numerical evaluation value \( \text{EvalValue} \). This value determines the relative ordering of the plaintexts and is compared against thresholds to classify the relationship. The classification results in one of two outcomes:
\begin{itemize}
    \item \textbf{True}: Indicates that \( m_a > m_b \), meaning \( m_a \) is ranked higher than \( m_b \) in the encrypted space.
    \item \textbf{False}: Indicates that \( m_a < m_b \), meaning \( m_b \) is ranked higher than \( m_a \).
\end{itemize}

This design avoids exposing explicit equality relationships (\( m_a = m_b \)) while preserving the total order of the encrypted values. Even if \( m_a \) and \( m_b \) are equal, the perturbation ensures that the ciphertexts corresponding to these plaintexts differ, preventing frequency analysis or other inference attacks.

The Perturbation-Aware Symbol Comparison algorithm ensures robust and privacy-preserving comparisons with minimal computational overhead. Its integration into the HADES framework supports privacy-preserving operations such as encrypted sorting, secure filtering, and range queries. By combining rigorous noise management with efficient polynomial arithmetic, the algorithm achieves high accuracy and scalability, making it suitable for large-scale encrypted datasets in real-world deployments.

\subsection{Correctness Analysis}

The correctness of the extended HADES scheme ensures that comparison operations accurately reflect the intended ordering of plaintexts, while equality relationships remain obfuscated. This is achieved through precise management of scaling, perturbations, and noise contributions.

Correctness is defined as the preservation of the relative ranking between plaintexts. For plaintexts \( m_a \) and \( m_b \), let:
\[
\text{TrueValue} = (m_a - m_b) \cdot \text{scale}
\]
represent the primary difference after scaling, and let:
\[
\text{PerturbationEffect} = (\Delta(m_a) - \Delta(m_b)) \cdot \text{scale}
\]
capture the perturbation contributions. Correctness holds if:
\[
|\text{PerturbationEffect}| \ll |\text{TrueValue}|,
\]
ensuring that the perturbation does not obscure the primary difference. This condition is satisfied when the perturbation range \( \Delta(m) \) is designed such that:
\[
|\Delta(m)| \ll |m_a - m_b|,
\]
making the perturbation negligible compared to the scaled plaintext difference when \( m_a \neq m_b \).

\paragraph{Noise Management}
Noise introduced during encryption and evaluation is bounded to ensure correctness. Let the noise term introduced by ciphertext operations be denoted as \( \langle e_m, ct_{\Delta} \rangle \). Correctness requires:
\[
|\langle e_m, ct_{\Delta} \rangle| < \frac{1}{2} |\text{TrueValue}|,
\]
ensuring that noise does not alter the sign of \( \text{EvalValue} \). By maintaining this constraint, the scheme guarantees that the scaled plaintext difference \( \text{TrueValue} \) dominates both perturbation and noise, preserving the accuracy of comparisons.


\paragraph{Scalability to Large Datasets}  
The HADES framework is designed to scale efficiently for large datasets by maintaining a fixed ciphertext size and leveraging the linear complexity of comparison operations. For \( n \) encrypted values, the comparison time scales as:
\[
T_{\text{cmp}} = \mathcal{O}(n),
\]
allowing HADES to handle datasets with millions of entries without significant performance degradation. Additionally, the linear complexity ensures that the framework can support high-throughput applications such as secure range queries and sorting.

\paragraph{Practical Implications}
By carefully balancing scaling, perturbations, and noise management, the HADES framework achieves accurate and robust comparisons. These properties ensure its applicability to real-world scenarios requiring secure and efficient computation, such as encrypted database queries, privacy-preserving analytics, and large-scale outsourced computations.

\subsection{Parameter Sensitivity}

The choice of key parameters, such as \( \text{scale} \) and \( \epsilon \), directly impacts the correctness, security, and efficiency of the HADES framework. Below, we analyze their effects.

\paragraph{Impact of Scaling Factor \( \text{scale} \)}  
A larger \( \text{scale} \) amplifies plaintext differences, making it easier to distinguish small perturbations and noise terms. However, excessively large \( \text{scale} \) increases computational overhead during encryption and evaluation. For practical applications, \( \text{scale} \) values in the range \( [10^2, 10^4] \) have shown to balance efficiency and robustness.

\paragraph{Impact of Perturbation Range \( \epsilon \)}  
The perturbation range \( \epsilon \) determines the degree of obfuscation for equality relationships. Larger \( \epsilon \) values provide stronger protection against frequency analysis but may reduce comparison correctness. Empirically, \( \epsilon \) values in \( [10^{-3}, 10^{-2}] \) are effective, ensuring that perturbation effects remain insignificant compared to scaled plaintext differences.

\subsection{Security Analysis}

The extended HADES scheme enhances security against frequency analysis attacks by introducing independent perturbations during encryption, while maintaining the core IND-CPA guarantees of the basic HADES scheme. A formal proof of IND-CPA security is provided, encompassing both the basic and extended algorithms.

\begin{theorem}
The extended HADES scheme is IND-CPA secure under the assumption that the RLWE problem is hard.
\end{theorem}

\begin{proof}
The IND-CPA security of the extended HADES scheme is proved via a reduction to the RLWE problem. Suppose an adversary \( \mathcal{A} \) can distinguish ciphertexts under the CPA model with non-negligible probability. We construct a reduction \( \mathcal{B} \) that uses \( \mathcal{A} \) to solve the RLWE problem.

\paragraph{Reduction Setup}
The RLWE challenge provides a tuple \( (a, b) \), where:
\[
b = a \cdot s + e \pmod{q}
\]
with \( s \) being the secret key and \( e \) a small noise polynomial, or \( b \) is uniformly random. The reduction uses \( (a, b) \) to simulate the encryption process as follows:
\begin{itemize}
    \item \( a \) is treated as the public key \( pk \).
    \item \( b \) is used to simulate the CEK by setting:
    \[
    cek = b.
    \]
\end{itemize}

\paragraph{Simulation of Encryption}
For plaintexts \( m_0 \) and \( m_1 \) submitted by \( \mathcal{A} \), the reduction selects a random bit \( b \in \{0, 1\} \), and encrypts \( m_b \) as:
\[
ct_b = \text{Encrypt}(pk, m_b \cdot \text{scale} + \Delta(m_b) \cdot \text{scale} + e_m \mod q),
\]
where \( \Delta(m_b) \) and \( e_m \) are generated according to the perturbation-aware encryption algorithm. The ciphertext \( ct_b \) is returned to \( \mathcal{A} \), simulating the encryption oracle in the IND-CPA game.

\paragraph{Distinguishing Capability}
If \( b \) follows the RLWE distribution, the ciphertexts \( ct_0 \) and \( ct_1 \) exhibit statistical properties consistent with the HADES encryption process. However, if \( b \) is uniformly random, the ciphertexts \( ct_0 \) and \( ct_1 \) become indistinguishable from random noise. The adversary \( \mathcal{A} \)'s ability to distinguish between these cases implies a solution to the RLWE problem.

\paragraph{Conclusion}
Since breaking the IND-CPA security of HADES implies solving the RLWE problem, the extended scheme is IND-CPA secure under the hardness of RLWE.
\end{proof}

\paragraph{Perturbation-Induced Randomness}
The inclusion of dynamic perturbations \( \Delta(m) \) ensures that ciphertexts corresponding to identical plaintexts are statistically independent. Specifically, for plaintexts \( m_a = m_b \), the perturbations \( \Delta(m_a) \) and \( \Delta(m_b) \) differ, producing distinct ciphertexts \( ct_{m_a} \neq ct_{m_b} \). This randomness eliminates the possibility of correlating ciphertexts based on plaintext equality, providing robustness against frequency analysis attacks.

\paragraph{Scaling and Noise Obfuscation}
The scaling factor \( \text{scale} \) amplifies plaintext differences, ensuring that the primary difference \( (m_0 - m_1) \cdot \text{scale} \) dominates perturbation and noise terms. Additionally, the bounded noise \( e_m \) ensures that ciphertexts remain computationally indistinguishable under the RLWE assumption. This dual-layer obfuscation prevents adversaries from inferring plaintext relationships, even when querying the same plaintext multiple times.

\paragraph{Enhanced Protection Against Equality Leakage}
Equality relationships (\( m_a = m_b \)) are explicitly obfuscated by combining perturbations and noise. Even with repeated queries, the adversary observes randomized outputs, making it infeasible to infer equality through statistical analysis or ciphertext patterns.

\paragraph{Scalability and Computational Efficiency}
Despite the added perturbations, the extended scheme introduces minimal computational overhead. The encryption and comparison processes retain their linear complexity with respect to the number of plaintexts, ensuring scalability for large datasets. By maintaining fixed ciphertext sizes and leveraging efficient polynomial arithmetic, the extended HADES scheme remains practical for real-world applications, such as encrypted database queries and privacy-preserving analytics.

\section{Evaluation}

\subsection{System Implementation}

The HADES framework was implemented using the OpenFHE library, a state-of-the-art homomorphic encryption framework that provides comprehensive support for multiple encryption schemes, including BFV and CKKS. The system integrates both the Basic and FA-Extension (FAE) functionalities under these schemes, enabling secure and efficient symbol comparisons in privacy-preserving applications. The source code for HADES will be hosted at
\[
\texttt{\url{https://github.com/hpdic/hades}}.
\]

\paragraph{Implementation of Basic HADES}  
The implementation of basic HADES provides fundamental functionality for secure symbol comparison. It includes:
\begin{itemize}
    \item \textbf{Key Generation (KeyGen)}: A public-private key pair is generated for encryption and decryption using OpenFHE's key generation routines. For BFV, a plaintext modulus of \( 65537 \) was used, while for CKKS, scaling modulus sizes and flexible scaling techniques were employed.
    \item \textbf{Encryption (EncBasic)}: Plaintext values are packed and encrypted using OpenFHE's encoding mechanisms. For BFV, plaintexts are directly encoded as integers, while for CKKS, floating-point values are transformed into complex numbers.
    \item \textbf{Comparison (CmpBasic)}: Ciphertext subtraction and evaluation are performed to compute the difference between two encrypted values. The result is decrypted and interpreted to determine the relative ordering.
\end{itemize}

\paragraph{FA-Extension (FAE) Implementation}  
The FA-Extension introduces perturbation-aware encryption to enhance security under stronger adversarial models:
\begin{itemize}
    \item \textbf{Perturbation Sampling}: During encryption, a scaling factor and a small random perturbation value (\(\Delta_m\)) are sampled. These perturbations obfuscate equality relationships between plaintext values, ensuring resistance against frequency analysis attacks.
    \item \textbf{Modified Encryption (EncFAE)}: Perturbed plaintexts are created by adding scaled perturbations to the original values. These perturbed plaintexts are then encrypted, following the same routines as in the Basic implementation.
    \item \textbf{Comparison (CmpFAE)}: The comparison semantics are enhanced to enforce strict unidirectional constraints, where equality obfuscation ensures that adversaries cannot deduce whether \(a = b\) by querying \(a \geq b\) and \(b \geq a\) simultaneously.
\end{itemize}

\paragraph{Encryption Schemes: BFV and CKKS}  
Both BFV and CKKS encryption schemes were implemented to support different computational requirements:
\begin{itemize}
    \item \textbf{BFV Scheme}: Designed for exact computations over integers, BFV leverages plaintext modulus \( 65537 \) and a multiplicative depth of \( 2 \) to enable efficient homomorphic additions and multiplications. Relinearization and rotation keys are precomputed for advanced operations.
    \item \textbf{CKKS Scheme}: CKKS supports approximate arithmetic for floating-point numbers, employing flexible scaling techniques to balance precision and performance. Key parameters include scaling modulus size \( 59 \), a ring dimension of \( 16384 \), and multiplicative depth \( 6 \).
\end{itemize}

\paragraph{OpenFHE Parameters and Optimizations}  
The implementation relies on various OpenFHE configurations to optimize performance and security:
\begin{itemize}
    \item \textbf{Security Levels}: For both BFV and CKKS schemes, the HEStd\_128\_classic security standard was adopted, ensuring 128-bit classical security.
    \item \textbf{Ring Dimensions}: BFV used a ring dimension of \( 4096 \), while CKKS employed \( 16384 \) for compatibility with approximate arithmetic.
    \item \textbf{Scaling Techniques}: CKKS utilized flexible and fixed auto-scaling techniques to minimize precision loss during computations.
    \item \textbf{Key Switching and Bootstrapping}: Advanced key switching was employed to optimize ciphertext manipulations. While BFV did not require bootstrapping, CKKS supported it to extend the operational depth for computationally intensive tasks.
\end{itemize}

\paragraph{System Compatibility and Scalability}  
The implementation is designed for scalability, supporting datasets with up to 35,848 data points. This limit ensures computational feasibility while maintaining generalizability. The framework can be extended to accommodate larger datasets by incorporating distributed encryption and parallelized comparison operations.

\subsection{Experimental Setup}

All experiments were conducted on a high-performance computing system~\cite{keahey2020lessons} with the following configuration:
\begin{itemize}
    \item \textbf{Processor}: Intel Xeon Gold 6248R CPU @ 3.00GHz, 48 cores.
    \item \textbf{Memory}: 256 GB DDR4 RAM.
    \item \textbf{Operating System}: Ubuntu 24.04 LTS.
    \item \textbf{Library}: OpenFHE version 1.2.3, compiled with GCC 10.3.0.
\end{itemize}

Each experiment was executed three times to account for runtime variations due to system load or random noise introduced by the FAE mechanism. The reported results represent the average of these three runs. Timing measurements were taken for each operation individually: Key Generation, Encryption (Basic and FAE), and Comparison (Basic and FAE). The average per-operation time was calculated and presented in milliseconds for ease of interpretation.

\subsubsection{Data Sets}

We evaluated the HADES framework on three real-world datasets, representing diverse application domains:
\begin{itemize}
    \item \textbf{Bitcoin}~\cite{bitcoin_trade}: A dataset containing 1,085 cryptocurrency transaction values.
    \item \textbf{Covid19}~\cite{covid19data}: A dataset with numeric metrics from Covid19-related data, 340 including case counts and recovery rates.
    \item \textbf{hg38}~\cite{hg_data}: A dataset containing 34,423 genomic data tuples in numeric form derived from human genome assembly \#38.
\end{itemize}

All of the above 35,848 values were preprocessed to fit the plaintext modulus constraints of the BFV scheme (\(65537\)) or appropriately scaled for the CKKS scheme. This ensures compatibility with encryption operations and avoids runtime errors.

\subsubsection{Baseline Protocols}

To contextualize our evaluation, we compare HADES against two prominent baseline schemes: HOPE~\cite{hope} and POPE~\cite{droche_ccs16}, which represent two distinct paradigms for ciphertext comparison in privacy-preserving outsourced databases.

HOPE is a stateless, cryptographically efficient scheme proposed in \cite{hope}. It leverages the Paillier~\cite{ppail_eurocrypt99} encryption system to achieve secure order-preserving comparisons without requiring server state or interaction between client and server during comparison. The primary advantage of HOPE lies in its efficiency, as it avoids the complexities of polynomial-based operations inherent in homomorphic encryption schemes like BFV and CKKS. However, HOPE supports only additive operations and is limited to integer-based computations. These restrictions make it unsuitable for applications requiring multiplicative homomorphic operations or floating-point arithmetic, reducing its applicability in real-world database systems.

POPE, introduced by Roche et al. \cite{droche_ccs16}, takes a client-dependent approach to ciphertext comparison. It encodes partial order information into the ciphertext, allowing the server to perform comparisons while preserving certain security guarantees. However, the scheme requires active client participation during comparison operations, introducing significant computational and communication overhead. This reliance on client involvement makes POPE less suitable for applications requiring fully independent server-side computation. Furthermore, the added network latency contributes to its inefficiency, particularly when compared to stateless schemes like HOPE or HADES.

\subsection{Micro Benchmarks}

To evaluate the performance of the HADES framework under controlled conditions, we conducted micro benchmarks using randomly generated datasets. Each dataset consisted of 100 numerical values, sampled uniformly at random within the range of \(0\) to \(10^6\). This approach ensures consistency and avoids biases introduced by specific real-world datasets.

The experiments measured the performance of key operations in both the Basic and FA-Extension (FAE) implementations. These operations included:
\begin{itemize}
    \item \textbf{Key Generation (KeyGen)}: Measuring the time to generate public-private key pairs and associated evaluation keys.
    \item \textbf{Basic Encryption (EncBasic)}: Encrypting 100 random values using the Basic implementation.
    \item \textbf{FA-Extension Encryption (EncFAE)}: Encrypting 100 random values with additional perturbation applied during the encryption process.
    \item \textbf{Basic Comparison (CmpBasic)}: Performing pairwise comparisons between ciphertexts to evaluate relative ordering.
    \item \textbf{FA-Extension Comparison (CmpFAE)}: Performing pairwise comparisons with the enhanced obfuscation of equality relationships.
\end{itemize}

\subsubsection{HADES for BFV}

\begin{figure}[t!]
    \centering
    \includegraphics[width=\linewidth]{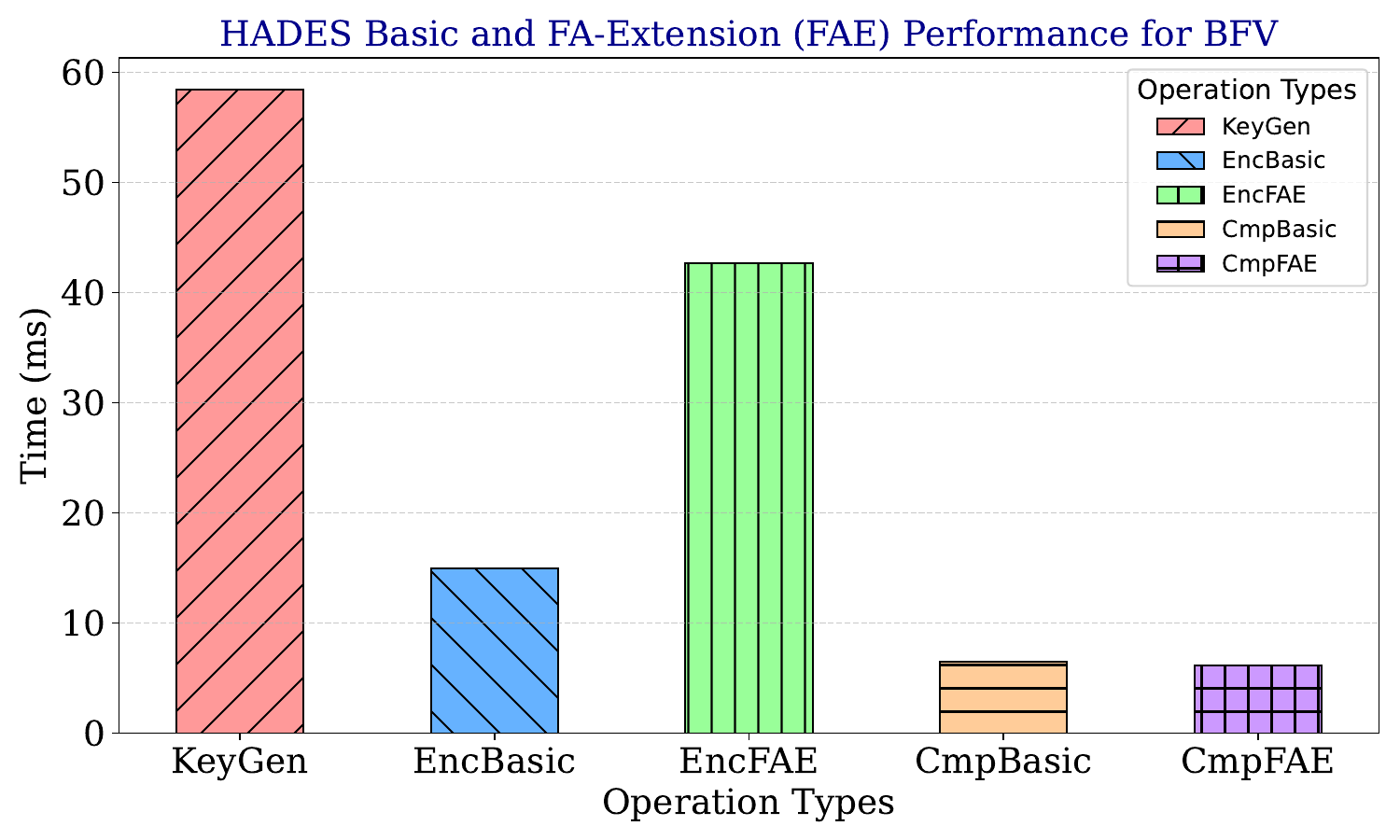}
    \caption{HADES Basic and FA-Extension (FAE) Performance for BFV~\cite{bfv}.}
    \label{fig:hades_bfv_performance}
\end{figure}

Figure~\ref{fig:hades_bfv_performance} presents the performance of the HADES framework implemented using the BFV encryption scheme. The analysis includes two key components: (1) the original HADES Basic scheme, and (2) its extension with perturbation-aware encryption, referred to as FAE (FA-Extension). The performance metrics evaluated are Key Generation Time, Average Encryption Time, and Average Comparison Time. The encryption and comparison times are presented separately for HADES Basic and FAE to highlight their differences.

The key generation time is consistent across both HADES Basic and FAE since it is independent of the encryption scheme;
therefore, we only report the performance once. 
For encryption, the FAE extension introduces additional steps, including perturbation sampling and noise addition, which increase the encryption time significantly compared to HADES Basic. Specifically, FAE encryption takes approximately three times longer than Basic encryption. 

On the other hand, comparison times are less affected by the extension since the perturbation does not significantly impact the operations required for ciphertext comparison. Interestingly, the comparison time for FAE is slightly lower than that of Basic, which could be due to the noise structure simplifying certain polynomial operations. These results demonstrate the trade-off between security enhancements and performance when adopting perturbation-aware encryption.

\subsubsection{HADES for CKKS}

\begin{figure}[t!]
    \centering
    \includegraphics[width=\linewidth]{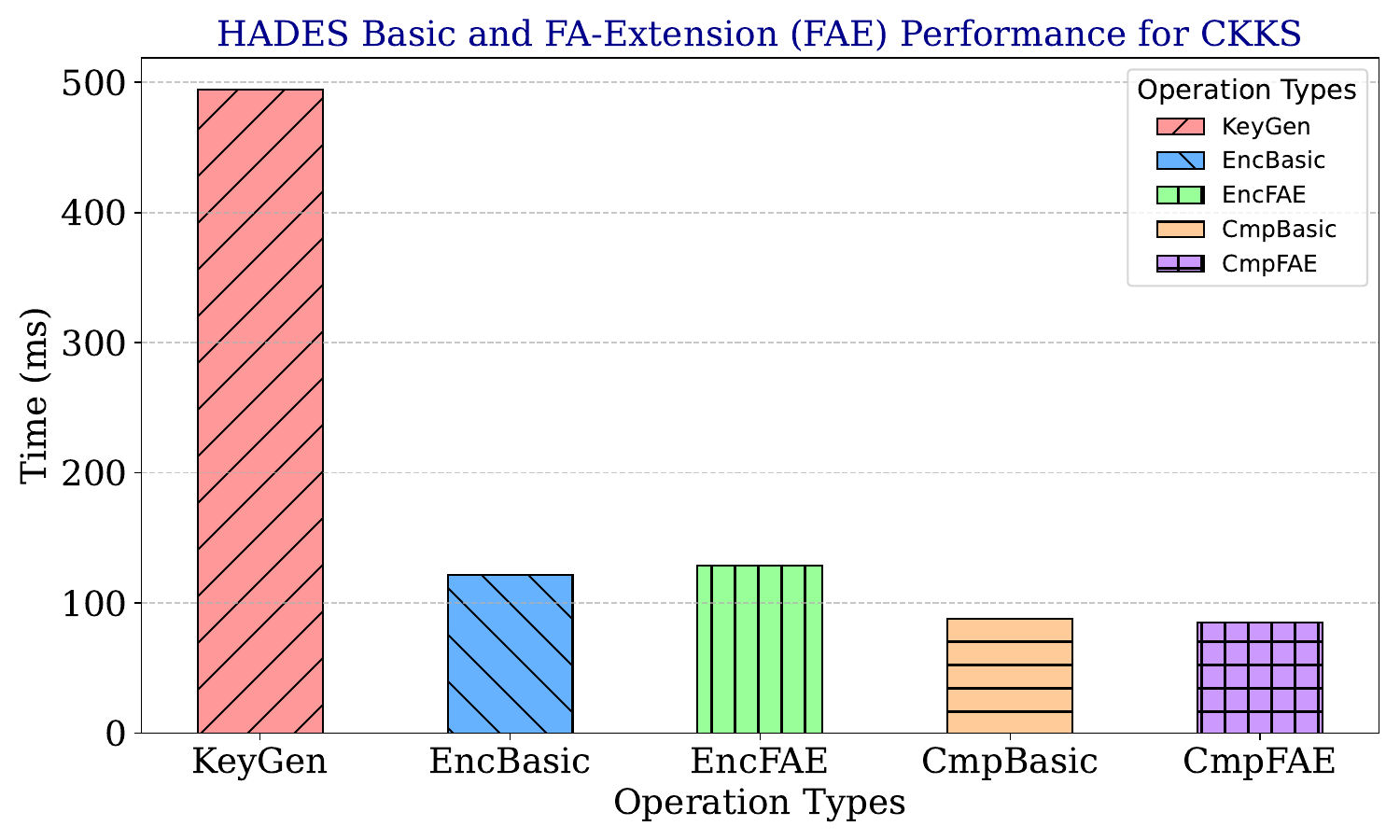}
    \caption{HADES Basic and FA-Extension (FAE) Performance for CKKS~\cite{ckks}.}
    \label{fig:hades_ckks_performance}
\end{figure}

Figure~\ref{fig:hades_ckks_performance} presents the performance analysis of the HADES framework using the CKKS encryption scheme. The analysis evaluates two configurations: (1) HADES Basic, and (2) HADES FA-Extension (FAE) with perturbation-aware encryption. The operations measured include Key Generation, Average Encryption Time, and Average Comparison Time, highlighting the differences between Basic and FA-Extension setups.

Key generation time remains consistent across all configurations as it is independent of specific encryption extensions. For encryption, the FA-Extension introduces additional perturbation steps, leading to a slight increase in encryption time compared to the Basic scheme. However, the comparison times between Basic and FA-Extension are almost identical, as the additional perturbation does not affect the comparison process significantly. These results demonstrate the effectiveness of CKKS in handling privacy-preserving computations while maintaining high precision.

Compared to BFV, CKKS demonstrates higher computation times across all operations. The increased time is due to CKKS's reliance on floating-point arithmetic and complex encoding, which require additional computational resources to maintain precision. For example, the encryption and comparison times in CKKS are approximately 2-3 times longer than their BFV counterparts. This trade-off is expected as CKKS offers enhanced flexibility and support for approximate arithmetic, making it more suitable for scenarios requiring floating-point operations.

\subsection{Real-world Datasets}

\begin{figure}[t!]
    \centering
    \includegraphics[width=\linewidth]{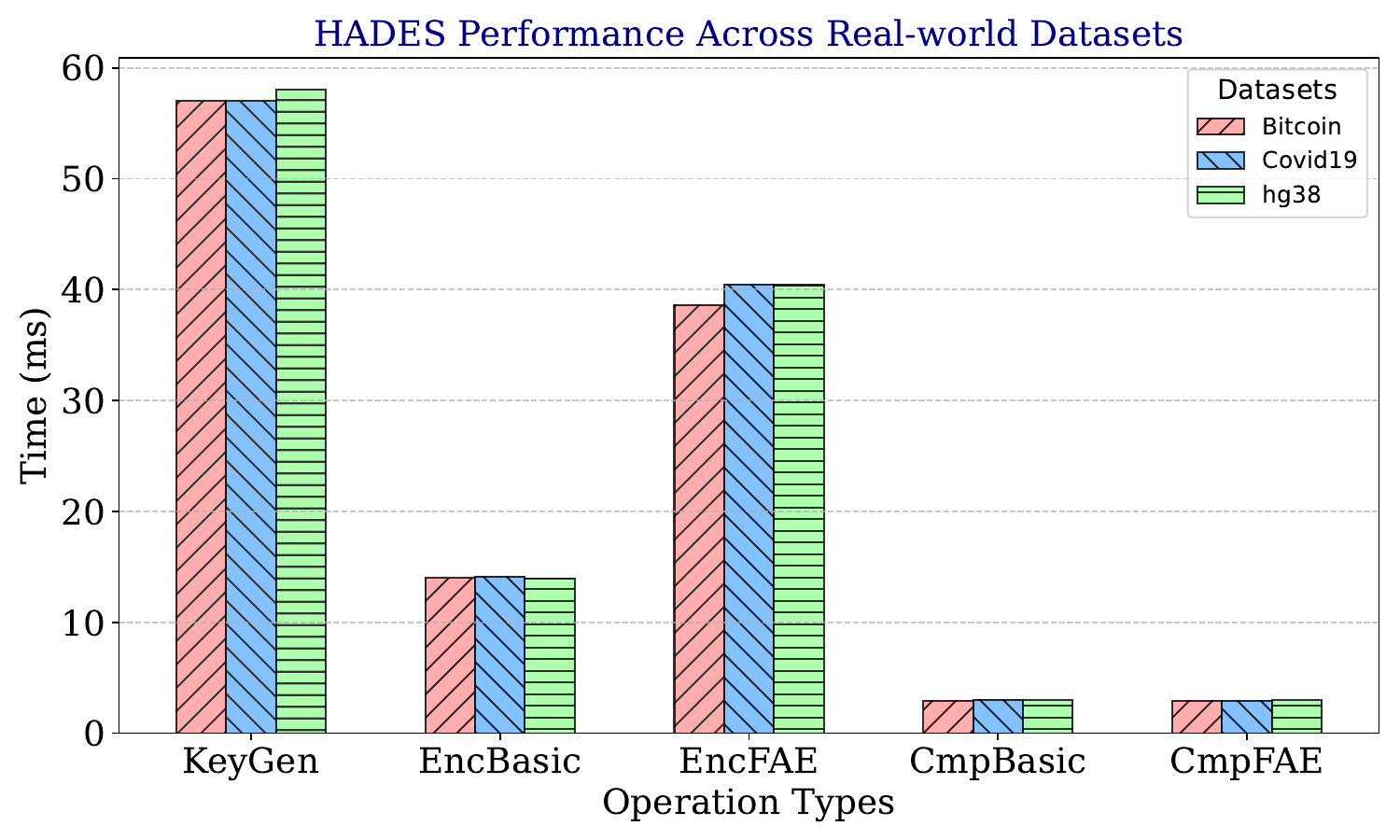}
    \caption{HADES Performance Across Datasets for Key Operations.}
    \label{fig:hades_app}
\end{figure}

Figure~\ref{fig:hades_app} illustrates the performance of the HADES framework implemented using the BFV encryption scheme, tested on three different datasets: Bitcoin, Covid19, and hg38. The performance evaluation measures five key operations: Key Generation (KeyGen), Basic Encryption (EncBasic), FA-Extension Encryption (EncFAE), Basic Comparison (CmpBasic), and FA-Extension Comparison (CmpFAE). Each operation's time is reported as an average per operation, with adjustments for comparability across datasets.

The reported times highlight the computational characteristics of the HADES BFV framework. KeyGen times are consistent across datasets, as they are independent of data characteristics. Encryption times (EncBasic and EncFAE) vary slightly due to the additional perturbation steps in FA-Extension, but they remain efficient relative to comparison times. Comparison times (CmpBasic and CmpFAE) dominate the overall computation, as pairwise operations scale quadratically with the dataset size. Notably, the measurements are reported per operation, ensuring comparability despite differences in data size and nature.

\subsection{Comparison to State-of-the-arts}

\begin{figure}[t!]
    \centering
    \includegraphics[width=\linewidth]{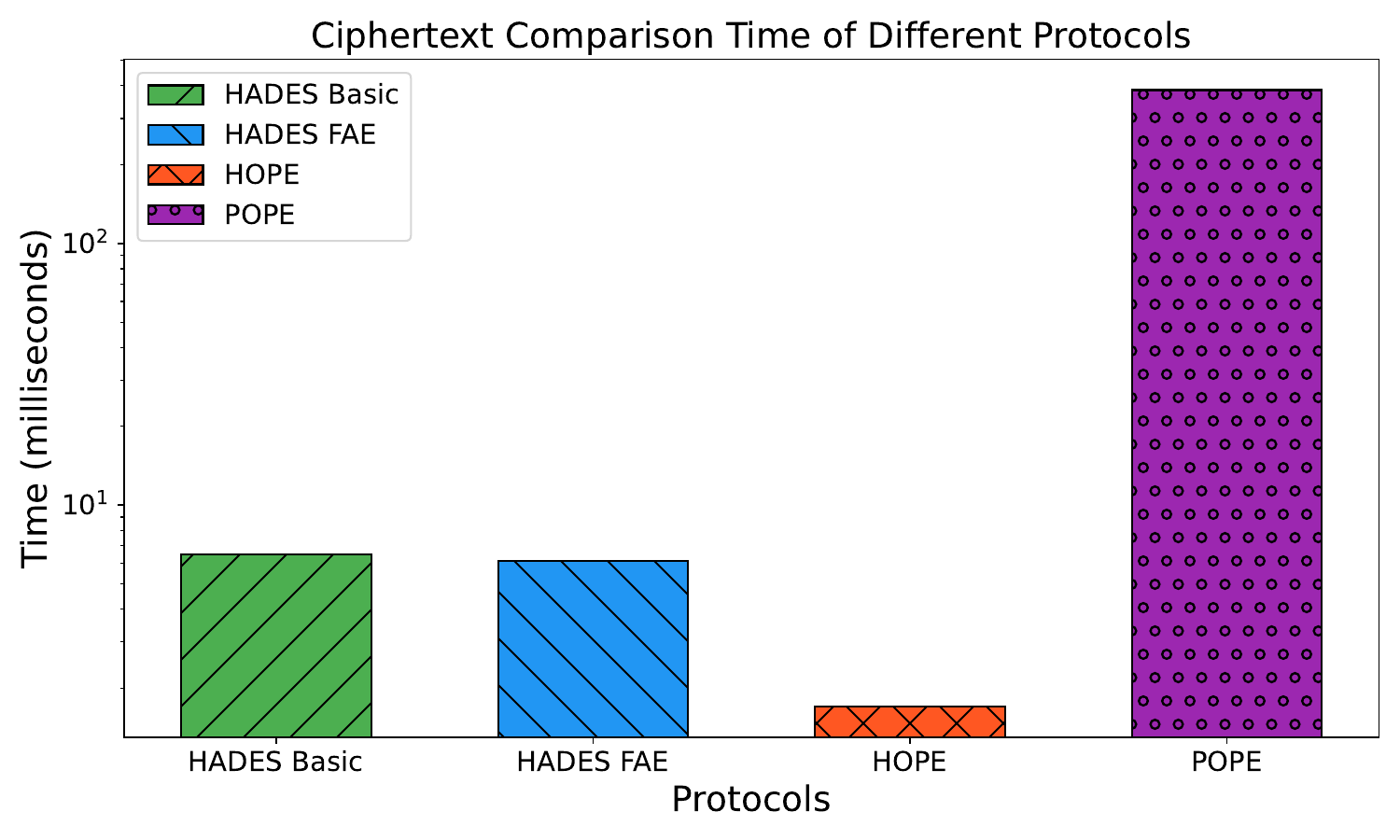}
    \caption{Ciphertext Comparison Time of Different Protocols.}
    \label{fig:comparison}
\end{figure}

We compared the ciphertext comparison time of our proposed HADES framework (both the Basic and FA-Extension variants) against state-of-the-art schemes, including HOPE~\cite{hope} and POPE~\cite{droche_ccs16}. The results, illustrated in Figure~\ref{fig:comparison}, demonstrate that HADES achieves competitive performance while ensuring stronger security guarantees. Specifically, the ciphertext comparison time for HADES Basic and HADES FA-Extension are 6.5ms and 6.1ms, respectively, which are comparable to HOPE (1.7ms) and significantly faster than POPE (385ms). 

The main reason for POPE's inefficiency lies in its reliance on client participation during comparison operations, which introduces not only computational overhead but also network latency. This architecture prevents fully independent ciphertext comparison on the server side, resulting in significant delays. HOPE, on the other hand, avoids these issues by being entirely stateless and cryptographically efficient. Built on Paillier encryption, it enables rapid comparisons but is limited by its reliance on an integer-based scheme that supports only addition. Additionally, HOPE's restricted functionality makes it less suitable for real-world database systems, which often require more advanced operations, such as multiplication or support for floating-point data.
In contrast, HADES strikes a balance between efficiency and functionality. While slightly slower than HOPE due to its reliance on polynomial-based homomorphic encryption, HADES offers support for both addition and multiplication, making it more suitable for modern database systems. Moreover, the extended FA-Extension variant ensures enhanced privacy protection against inference attacks, addressing critical shortcomings of existing stateless schemes like HOPE. These advantages make HADES a robust choice for privacy-preserving database applications.

\section{Conclusion}

This paper presents HADES, a novel cryptographic framework enabling efficient and secure symbol comparison within fully homomorphic encryption (FHE) without ciphertext expansion. HADES introduces the Compare-Eval Key (CEK) mechanism, rigorously proven CPA-secure under the Ring Learning with Errors (RLWE) problem. We provide a detailed theoretical analysis, including concrete parameter selection to achieve desired security levels. This framework supports accurate comparisons while maintaining computational and storage efficiency, addressing advanced threats such as frequency-analysis attacks through a novel perturbation-aware encryption technique. Implemented using OpenFHE, HADES demonstrates practical performance on real-world datasets, supporting both integer and floating-point data through BFV and CKKS schemes and outperforming state-of-the-art baselines in relevant benchmarks. These results establish HADES as a robust and scalable solution for privacy-preserving computations, particularly in outsourced database scenarios requiring efficient range queries and indexing. Future research directions include exploring optimizations for specific hardware architectures and investigating the applicability of HADES to other privacy-preserving computation paradigms.

\section*{Acknowledgment}
Results presented in this paper were partly obtained using the Chameleon testbed supported by the National Science Foundation. We also appreciate the insightful discussions and feedback from Professor Stefano Tessaro at the University of Washington, which helped refine some of the key ideas presented in this paper.

\bibliographystyle{ACM-Reference-Format}
\bibliography{references}

\end{document}